\documentclass[twocolumn,showpacs,preprintnumbers,amsmath,amssymb]{revtex4}
\usepackage{graphicx}% Include figure files
\usepackage{dcolumn}% Align table columns on decimal point
\usepackage{bm}% bold math
\usepackage{subfigure}
\usepackage{color}

\begin{document}

\title{Double parton scattering in double logarithm approximation of perturbative QCD}

\author{M.G.~Ryskin}
\affiliation{Petersburg Nuclear Physics Institute, NRC Kurchatov Institute, Gatchina, St. Petersburg, 188300, Russia}

\author{A.M.~Snigirev}
\affiliation{Skobeltsyn Institute of Nuclear Physics, Lomonosov Moscow State University, 119991, Moscow, Russia}

\date{\today}
\begin{abstract}
Using the explicit form of the known single distribution functions (the Green's functions) in the double logarithm approximation of perturbative QCD, we analyze the structure of splitting diagrams as a source of double parton perturbative correlations in the proton. The related phenomenological effects are discussed for the conditions of the LHC experiments.
\end{abstract}

%\setpagewiselinenumbers
%\linenumbers

%\keywords:{ Double parton scattering, QCD evolution, two-parton distributions}
\maketitle
\section{\label{sec1}Introduction}
Strong interest has arisen~\cite{Bartalini:2011jp}  in the investigations and measurements of the multiparton interactions in high energy hadron-hadron collisions.
The analysis of final states with four jets, $\gamma+3$ jets, and $W+2$ jets, performed by the AFS~\cite{AFS}, UA2~\cite{UA2}, CDF~\cite{cdf4jets,cdf}, D0~\cite{D0}, and ATLAS~\cite{atlas} Collaborations, provides convincing evidence for the significance of hard multiple parton interactions in these collisions  and thereby supplements our understanding of the proton structure with new information. Studies of hard double parton scattering (DPS) have a long history theoretically, with many references to prior work listed, for instance, in the recent review~\cite{Bartalini:2011jp}. A greater rate of events containing multiple hard interactions is anticipated at the LHC,  with respect to the experiments mentioned above, due to the much higher luminosity and greater energy of the LHC. Moreover, the DPS processes can constitute an important background~\cite{DelFabbro:2002pw,Hussein:2006xr} to signals from the Higgs and other interesting processes. Besides, certain types of multiple interactions will have distinctive signatures~\cite{Kulesza:1999zh,Cattaruzza:2005nu,maina,Berger:2009cm,Berger:2011ep}, facilitating 
a detailed investigation of these processes experimentally and revealing information about parton pair correlations in the proton.

The inclusive cross section of a DPS process in a hadron collision with the two hard parton subprocesses $A$ and $B$ may be written in the factorized form [see, for instance, Ref.~\cite{Diehl:2011yj}, where Eq.~(\ref{hardAB_p}) is derived in detail] as
\begin{eqnarray}
\label{hardAB_p}
\sigma^D_{(A,B)} =& & \frac{m}{2} \sum \limits_{i,j,k,l} \int \Gamma_{ij}(x_1, x_2; {\bf q}; Q^2_1, Q^2_2)\nonumber\\
& &\times \hat{\sigma}^A_{ik}(x_1, x_1^{'}) 
\hat{\sigma}^B_{jl}(x_2, x_2^{'}) \Gamma_{kl}(x_1^{'}, x_2^{'}; {\bf -q}; Q^2_1, Q^2_2)\nonumber\\ 
& & \times dx_1 dx_2 dx_1^{'} dx_2^{'} \frac{d^2q}{(2\pi)^2}.
\end{eqnarray}
Here $\Gamma_{ij}(x_1, x_2; {\bf q} ; Q^2_1, Q^2_2)$ are the generalized double parton distribution functions, depending on the longitudinal momentum fractions $x_1$ and $x_2$ of the two partons undergoing the hard processes $A$ and $B$ at the scales $Q_1$ and $Q_2$. $\hat{\sigma}^A_{ik}$ and $\hat{\sigma}^B_{jl}$ are the parton-level subprocess cross sections. The factor $m/2$ is a consequence of the symmetry of the expression for interchanging parton species $i$ and $j$. $m=1$ if $A=B$ and $m=2$ otherwise. Note that these distribution functions also depend on the transverse vector {\bf q} which is equal to the difference of the momenta of partons from the wave function of the colliding hadrons in the amplitude and the amplitude conjugated. Such dependence arises because the difference of parton transverse momenta within the parton pair is not conserved.
This transverse momentum {\bf q} is the Fourier conjugated variable of the parton pair transverse separation.
The starting cross section formula~(\ref{hardAB_p}) is somewhat similar to that usually used for single parton scattering. It was found (derived) in many works using the light-cone variables and the same approximations as those applied to the processes with a single hard scattering.

The main problem is to make the correct calculation of the two-parton functions $\Gamma_{ij}(x_1, x_2; {\bf q}; Q^2_1, Q^2_2)$ without~\cite{Diehl:2011yj,Ryskin:2011kk,Blok:2010ge,Diehl:2011tt,stir,flesburg} additional simplifying assumptions (the factorization of the impact-parameter dependence and $x$ dependence that, by no means, should be treated as inviolate). These functions and the corresponding evolution equations were considered in the current literature~\cite{Kirschner:1979im,Shelest:1982dg,snig03,snig04} only for ${\bf q}=0$ in the collinear approximation. In this approximation the two-parton distribution functions, 
$$\Gamma_{ij}(x_1, x_2; {\bf q}=0; Q^2, Q^2)=D^{ij}_h(x_1, x_2; Q^2, Q^2),$$
with the two hard scales set equal, satisfy the generalized Dokshitzer-Gribov-Lipatov-Altarelli-Parisi (DGLAP) evolution equations, derived initially in Refs.~\cite{Kirschner:1979im,Shelest:1982dg}. Likewise, the single distributions satisfy more widely known and often cited DGLAP equations~\cite{gribov,lipatov,dokshitzer,altarelli}. The functions in question have a specific interpretation in the leading logarithm approximation of perturbative QCD: they are the inclusive probabilities that in a hadron $h$ one finds two bare partons  of  types $i$ and $j$ with the given longitudinal momentum fractions $x_1$ and $x_2$.

Based on these well-known collinear  distributions, we have recently suggested~\cite{Ryskin:2011kk} a practical method which makes it possible to estimate the inclusive cross section for a DPS
process without  the oversimplified {\em additional} factorization assumption
for $\Gamma_{ij}(x_1,x_2; {\bf q}=0; Q^2_1,Q^2_2)=D^i_h(x_1;Q^2_1)D^j_h(x_2;Q^2_2)$
(which, in general, is inconsistent with the QCD evolution)
but taking into account the QCD evolution explicitly~\footnote{In $1\times
1$ and $1\times 2$ contributions of Eqs. (\ref{hardAB1}) and (\ref{D1xD2_s}) the factorization
$D_h^{ij}(x_1,x_2;\mu^2,\mu^2) = D_h^i(x_1;\mu^2) D_h^j(x_2;\mu^2)$ is assumed
at a low starting scale $\mu^2$ only.}. Afterwards, similar results were obtained also in Ref.~\cite{Blok:2011bu}, with an emphasis on the differential cross sections, and were partly supported in Ref.~\cite{Gaunt:2012wv}, albeit with some diversity of opinion regarding a terminology mainly. We found that single and double perturbative splitting graphs can meaningfully contribute to the inclusive cross section for a DPS process, in comparison with a ``traditional'' factorization component.

The main purpose of the present paper is to analytically study the structure of these single and double perturbative splitting diagrams as a source of parton pair perturbative correlations in the proton. The paper is organized as follows. In order to be clear and to introduce the denotations, we briefly recall some basic formulas from our previous work~\cite{Ryskin:2011kk} in Sec. II. The double parton correlations in the double logarithm approximation of perturbative QCD are estimated in Sec. III. The possible phenomenological issues at the LHC are discussed in Sec. IV, together with conclusions.

\section{\label{sec2}Inclusive cross section in terms of collinear distributions}
The inclusive cross section for DPS can be presented in the following form~\cite{Ryskin:2011kk}:
\begin{eqnarray}
\label{ABtotal}
\sigma^D_{(A,B)} =\sigma^{D,1\times1}_{(A,B)}+ \sigma^{D,1\times2}_{(A,B)}+\sigma^{D,2\times2}_{(A,B)},
\end{eqnarray}
where
\begin{widetext}
%where
\begin{eqnarray}
\label{hardAB1}
\sigma^{D,1\times1}_{(A,B)} = & &\frac{m}{2} \sum \limits_{i,j,k,l} \int D^i_h (x_1; \mu^2, Q^2_1) D^j_h (x_2;\mu^2, Q^2_2) 
\hat{\sigma}^A_{ik}(x_1, x_1^{'})
\hat{\sigma}^B_{jl}(x_2, x_2^{'}) 
D^k_{h'} (x'_1; \mu^2, Q^2_1)
D^l_{h'} (x'_2;\mu^2, Q^2_2)\nonumber\\
& &\times dx_1 dx_2 dx_1^{'} dx_2^{'} \int  F_{2g}^4(q)\frac{d^2q}{(2\pi)^2},
\end{eqnarray}
%\begin{widetext}
\begin{eqnarray}
\label{D2xD2}
\sigma^{D,2\times2}_{(A,B)} 
=& & \frac{m}{2} \sum \limits_{i,j,k,l} \int dx_1 dx_2 dx_1^{'} dx_2^{'} \int\limits^{\min(Q_1^2,Q_2^2)} \frac{d^2q}{(2\pi)^2}\nonumber\\
& &\times \sum\limits_{j{'}j_1{'}j_2{'}} \int\limits_{q^2}^{\min(Q_1^2,Q_2^2)}dk^2 \frac{\alpha_s(k^2)}{2\pi k^2}
\int\limits_{x_1}^{1-x_2}\frac{dz_1}{z_1}
\int\limits_{x_2}^{1-z_1}\frac{dz_2}{z_2}
 D_h^{j{'}}(z_1+z_2;\mu^2,k^2)\nonumber\\
& & \times \frac{1}{z_1+z_2}P_{j{'} \to
j_1{'}j_2{'}}\Bigg(\frac{z_1}{z_1+z_2}\Bigg)
D_{j_1{'}}^{i}(\frac{x_1}{z_1};k^2,Q_1^2) 
D_{j_2{'}}^{j}(\frac{x_2}{z_2};k^2,Q_2^2)
\hat{\sigma}^A_{ik}(x_1, x_1^{'}) \hat{\sigma}^B_{jl}(x_2, x_2^{'})\nonumber\\
& &\times\sum\limits_{j{'}j_1{'}j_2{'}} \int\limits_{q^2}^{\min(Q_1^2,Q_2^2)}dk^{'2} \frac{\alpha_s(k^{'2})}{2\pi k^{'2}}
\int\limits_{x'_1}^{1-x'_2}\frac{dz'_1}{z'_1}
\int\limits_{x'_2}^{1-z'_1}\frac{dz'_2}{z'_2}
D_{h'}^{j{'}}(z'_1+z'_2;\mu^2,k^{'2})\nonumber\\
& &\times \frac{1}{z'_1+z'_2}P_{j{'} \to
j_1{'}j_2{'}}\Bigg(\frac{z'_1}{z'_1+z'_2}\Bigg)
D_{j_1{'}}^{k}(\frac{x'_1}{z'_1};k^{'2},Q_1^2) 
D_{j_2{'}}^{l}(\frac{x'_2}{z'_2};k^{'2},Q_2^2),
\end{eqnarray}
or in substantially shorter yet less transparent form,
\begin{eqnarray}
\label{D2xD2_s}
\sigma^{D,2\times2}_{(A,B)} 
=& & \frac{m}{2} \sum \limits_{i,j,k,l} \int dx_1 dx_2 dx_1^{'} dx_2^{'} \int\limits^{\min(Q_1^2,Q_2^2)} \frac{d^2q}{(2\pi)^2}\nonumber\\
& &\times D_{h2}^{ij}(x_1,x_2;q^2, Q_1^2, Q_2^2)
\hat{\sigma}^A_{ik}(x_1, x_1^{'}) 
\hat{\sigma}^B_{jl}(x_2, x_2^{'})
D_{h'2}^{kl}(x'_1,x'_2;q^2, Q_1^2, Q_2^2),
\end{eqnarray}
and for the combined (``interference'') contribution,
\begin{eqnarray}
\label{D1xD2_s}
\sigma^{D,1\times2}_{(A,B)} 
=& & \frac{m}{2} \sum \limits_{i,j,k,l} \int dx_1 dx_2 dx_1^{'} dx_2^{'} \int\limits^{\min(Q_1^2,Q_2^2)} F_{2g}^2(q)\frac{d^2q}{(2\pi)^2}\nonumber\\
& &\times [D^i_h (x_1; \mu^2, Q^2_1) D^j_h (x_2;\mu^2, Q^2_2)
\hat{\sigma}^A_{ik}(x_1, x_1^{'}) 
\hat{\sigma}^B_{jl}(x_2, x_2^{'})
D_{h'2}^{kl}(x'_1,x'_2;q^2, Q_1^2, Q_2^2)\nonumber\\
& &+D_{h2}^{ij}(x_1,x_2;q^2, Q_1^2, Q_2^2)
\hat{\sigma}^A_{ik}(x_1, x_1^{'})
\hat{\sigma}^B_{jl}(x_2, x_2^{'})
D^k_{h'} (x'_1; \mu^2, Q^2_1) D^l_{h'} (x'_2;\mu^2, Q^2_2)].
\end{eqnarray}
\end{widetext}
Here $\alpha_s(k^2)$ is the QCD coupling, and 
$D_{j_1{'}}^{j_1}(z;k^2,Q^2)$ are the known single distribution functions (the Green's functions generated by the usual DGLAP kernels) at the parton level with the specific $\delta$-like  initial conditions at $Q^2=k^2$. The one-parton distribution (before splitting into the two branches at some scale $k^2$) is given by $D_h^{j'}(z_1+z_2;\mu^2,k^2)$.
The splitting functions
$$ \frac{1}{z_1+z_2}P_{j{'} \to j_1{'}j_2{'}}\Bigg(\frac{z_1}{z_1+z_2}\Bigg)$$
are the nonregularized one-loop well-known DGLAP kernels without the ``$ + $'' prescription. The single parton distribution functions $D^i_h (x_1; \mu^2, Q^2_1)$ are the solutions of the DGLAP equations with the given initial conditions $D^i_h (x_1; \mu^2)$ at the reference scale $\mu^2$ and may be expressed via the Green's functions $D^i_{i'}(z;k^2,Q^2)$ in the following way:
\begin{eqnarray}
\label{1solution}
& & D_h^i(x; \mu^2, Q^2)\nonumber\\
& & = \sum\limits_{i{'}} \int \limits_x^1
\frac{dz}{z}~D_h^{i{'}}(z;\mu^2)~D_{i{'}}^i(\frac{x}{z};\mu^2, Q^2).
\end{eqnarray}

Let us first consider the $1\times 1$ component which describes the two hard subprocesses $A$ and $B$ caused by the interactions of two pairs of partons in two independent branches of parton cascades. The probability of this double parton interaction depends on the spatial distribution (in the impact-parameter transverse plane) of these two branches of the parton cascade. In the momentum representation the spatial distribution is regulated by the two-parton (in a low $x$ region this is mainly the two-gluon) form factor $F_{2g}(q)$. After the integration over $q$ it gives the factor
 $$  \int  F_{2g}^4(q)\frac{d^2q}{(2\pi)^2}=\frac {1}
{\sigma_{\rm eff}}$$
which characterizes the transverse area occupied by the partons participating in  hard collision and is often denoted as an effective cross section, $\sigma_{\rm eff}$.

Thus the value of the DPS cross section depends on the spatial correlations between the two partons in the incoming proton (hadron) wave function. Because of a strong $k_\perp$ ordering during the DGLAP evolution, the position of the parton in the impact-parameter, $b_t$, plane is frozen, and the form factor $F_{2g}$ describes the initial $b_t$ distribution formed in the nonperturbative region somewhere at a low scale, less than $\mu^2$, where the DGLAP evolution starts. 

However, there is another sort of correlation caused by the splitting of one branch of the parton cascade into two branches. A splitting at the scale $k^2$  produces two branches placed rather 
close to each other, with the spatial separation $\delta b_t^2 \sim 1/k^2$. This effect is of perturbative origin and it may noticeably enlarge the  DPS cross section, leading to a lower mean value of $\sigma_{\rm eff}$.

Depending on the kinematics of the DPS process and experimental cuts, we may concentrate on the two-parton correlations coming from the nonperturbative region, that is, on the
$1\times 1$ contribution or on the correlations of perturbative origin, like that in the $2\times 2$ term.

The contribution of the combined component ($1\times2$) is regulated
by the  form factor $F_{2g}(q)$~\cite{Blok:2010ge,Frankfurt:2002ka} from the side of one incoming proton and  by the perturbative splitting on another side. Since the form factor $F_{2g}(q)$ already provides the convergence of the $q$ integral in the low $q^2<\mu^2$ region, here we mainly deal with the ``long distance'' correlations from the nonperturbative region.

The nonperturbative correlations corresponding to a low $q^2$ were discussed in detail in Refs. \cite{Blok:2010ge,Blok:2011bu},  where it was proposed to consider the DPS events with  small transverse momenta of the systems $A$ and $B$ produced by the hard subprocesses. This would be a very interesting study, albeit such  cuts select a very small part of the total DPS cross section.
 An alternative possibility to study the correlations at a low 
 scale $q^2$ is to measure the asymmetric DPS processes where one 
 ``hard'' scale $Q^2_1$ is relatively low, say, the 
$\chi_c$-meson and high-$E_T$ dijet DPS processes. Since  $q^2<Q^2_1$ ($Q_1$ is a relatively low scale corresponding to $\chi_c$ production) we  have practically
no space for pQCD splitting, and the $1\times 1$ configuration will dominate.

There is some discussion in the current literature~\cite{stir,Blok:2011bu,Gaunt:2012wv,Manohar} concerning the $2\times 2$ component. This contribution incorporates the  two splitting functions and the integration over $q$ without the strong suppression factor $F_{2g}(q)$. Formally, in the region of not too small $x$, within the
collinear approach this contribution should be considered as a result of the interaction of {\it one} pair of partons with the 
$2\to 4$ hard subprocess~\cite{stir,Blok:2011bu,Gaunt:2012wv,Manohar}, since  the dominant contribution to the phase space integral comes from a large $q^2\sim \min(Q_1^2,Q_2^2) $. However, as we argue in Ref.~\cite{Ryskin:2011kk} there may be configurations with a rather large interval between the splitting point $(k^2,\, z_1+z_2$) and the (momentum) coordinates $(Q^2_i,\, x_i)$ of the hard subprocess. In such a case the interval will be filled by the evolution  [either DGLAP when $k^2<<Q^2_i$ or Balitsky-Fadin-Kuraev-Lipatov  (BFKL)~\cite{ryskin,bfkl,bfkl2,bal} when $z_1+z_2>>x_i$] which will produce  additional secondaries. The corresponding process can not be described by the $2\to 4$ hard matrix element.  Here we have to use our formulas~(\ref{ABtotal}) and (\ref{D2xD2}), especially in the case of a configuration with two quite different scales (for instance, $Q^2_1<<Q^2_2$). 

In order to understand the structure of these additional contributions ($1\times2$ and $2\times2$) better, we
consider them in the double logarithm approximation, which allows us to obtain some analytical estimations.

\section{\label{sec3}Double logarithm approximation}
Let us write down all the integrations with splitting functions separately to make the analysis more transparent,
\begin{eqnarray}
\label{splitting}
& & D^{ij}_{h2}(x_1,x_2;q^2,Q_1^2,Q_2^2)\nonumber\\
= & &\sum\limits_{j{'}j_1{'}j_2{'}} \int\limits_{q^2}^{\min(Q_1^2,Q_2^2)}dk^2 \frac{\alpha_s(k^2)}{2\pi k^2}
\int\limits_{x_1}^{1-x_2}\frac{dz_1}{z_1}
\int\limits_{x_2}^{1-z_1}\frac{dz_2}{z_2}\nonumber\\
& &\times D_h^{j{'}}(z_1+z_2;\mu^2,k^2)
\frac{1}{z_1+z_2}P_{j{'} \to
j_1{'}j_2{'}}\Bigg(\frac{z_1}{z_1+z_2}\Bigg)\nonumber\\
& & \times D_{j_1{'}}^{i}(\frac{x_1}{z_1};k^2,Q_1^2) 
D_{j_2{'}}^{j}(\frac{x_2}{z_2};k^2,Q_2^2).
\end{eqnarray}

Because of the strong suppression factor $F_{2g}^2(q)$ in a single splitting diagram ($1\times2$ contribution), this integral $D^{ij}_{h2}(x_1,x_2;q^2,Q_1^2,Q_2^2)$ can be estimated at the reference scale $q^2=\mu^2$  
%in single splitting diagrams ($1\times2$ contribution)
and can be considered as the QCD evolution correction to the factorized double parton distribution functions. For double splitting diagrams ($2\times2$ contribution) we should keep the further {\it nonlogarithmic} integration over $q$ in mind.

In the double logarithm approximation we can restrict ourselves to the main gluon contribution only  and rewrite the integral under consideration in the following form~\footnote{After changing  the integration variables: $z_1=uz$, $z_2=u(1-z)$.}:
\begin{eqnarray}
\label{splitting-gluon}
& & D^{gg}_{h2}(x_1,x_2;q^2,Q_1^2,Q_2^2)\nonumber\\
& &= \int\limits_{q^2}^{\min(Q_1^2,Q_2^2)}dk^2 \frac{\alpha_s(k^2)}{2\pi k^2}
\int\limits\frac{du}{u^2}
 D_h^{g}(u;\mu^2,k^2)\int\limits \frac{dz}{z(1-z)}\nonumber\\
& & \times P_{g \to gg}(z)
D_{g}^{g}(\frac{x_1}{uz};k^2,Q_1^2) 
D_{g}^{g}\left(\frac{x_2}{u(1-z)};k^2,Q_2^2\right),
\end{eqnarray}
where $u=z_1+z_2$ and $z=z_1/u$.

The limits in the $u$ and $z$ integrations are  $x_1< uz$, $x_2<u(1-z)$, $u<1$, and $z<1$. The Green's functions (gluon distributions at the parton level) in the double logarithm approximation (see, for instance, \cite{dokshitzer,ryskin}) read
\begin{eqnarray}
\label{mellin in x=0}
 x D_g^{g}(x,t)
\simeq 4N_ct v^{-3/2}\exp{[v-at]}/\sqrt{2\pi},
\end{eqnarray}
where $v=\sqrt{8N_ct\ln{(1/x)}}$, $a=\frac{11}{6}N_c +\frac{1}{3}n_f/N^2_c$,
\begin{equation}
\label{tq}
t(Q^2) = \frac{2}{\beta}\ln\Bigg
[\frac{\ln(\frac{Q^2}{\Lambda^2} )}
{\ln(\frac{\mu^2}{\Lambda^2})}\Bigg],
\end{equation}
and where $\beta = (11N_c-2n_f)/3$, with the number of active flavors $n_f$, $\Lambda$ is the dimensional QCD parameter, and $N_c=3$ is the color number. Recall that in Eq.~(\ref{tq}) the one-loop running QCD coupling 
\begin{equation}
\label{alp}
\alpha_s(Q^2)=\frac{4\pi}{\beta\ln(Q^2/\Lambda^2)}
\end{equation}
was used. After that, the integral
(\ref{splitting-gluon}) may be rewritten as
\begin{eqnarray}
\label{splitting-gluon-t} 
& & x_1 x_2 D^{gg}_{h2}(x_1,x_2;\tau,T_1,T_2)\nonumber\\
\sim & & \int\limits_{\tau}^{\min(T_1,T_2)}dt
%\int\limits dy 
\int\limits dz
P_{g \to gg}(z)\int dy\nonumber\\
& &\times \exp{[\sqrt{8N_c} d(t,y,z)]},
\end{eqnarray}
where
\begin{eqnarray}
 d(t,y,z) = & &\sqrt{ty}
+ \sqrt{(T_1-t)(Y_1-y)}\nonumber\\
& & +\sqrt{(T_2-t)(Y_2-y)}
\end{eqnarray}
with $t=t(k^2)$, $T_1=t(Q_1^2)$, $T_2=t(Q_2^2)$, $\tau=t(q^2)$;
and $y=\ln(1/u)$, $Y_1=\ln(1/x_1)-\ln(1/z)$, $Y_2=\ln(1/x_2)-\ln(1/(1-z))$.

In Eq.~(\ref{splitting-gluon-t}) we keep the leading exponential terms only, which have the same structure both at the parton level and at the hadron level under smooth enough initial conditions at the reference scale. Indeed, in the double logarithm approximation Eq. (\ref{1solution}) reads 
\begin{eqnarray}
\label{1gsolution}
 x D_h^g(x; T) & & \simeq
\int \limits_0^Y dy'[z'D_h^{g}(z',0)]|_{1/z'=\exp{y'}}\nonumber\\
& & \times \exp{[\sqrt{8N_c}\sqrt{T(Y-y')} ]}\nonumber\\
 & & \sim \exp{[\sqrt{8N_c}\sqrt{TY} ]}
\end{eqnarray}
with $T=t(Q^2)$ and $Y=\ln(1/x)$.
The $y'$ integration is not a saddle-point type, and therefore, one of the edges, namely $y'\to 0$ ($z'\to 1$), dominates, provided that the initial gluon distribution does not grow too much as $z'$ decreases. In fact, one needs  $z'D_h^{g}(z',0) \sim (1/z')^a$  at $z' \to 0$ with $a<A$, where $A=\sqrt{2N_cT/Y} >0$. Note that the parametrization of the initial gluon distributions usually used satisfies this condition (e.g., the CTEQ parametrization from Ref. ~\cite{cteq}).

We are interested in the domain with large enough $T_1$, $T_2$, $\ln(1/x_1)$, and $\ln(1/x_2)$, when the exponential factors are large in comparison with 1 and where the approximations above are justified. In this case the integration over  the rapidity $y$ has a saddle-point structure in the wide interval of $z$ integration not near the kinematic boundaries. 
The saddle-point equation reads
\begin{eqnarray}
 \frac{\sqrt{t}}{\sqrt{y_0}}
-\frac{\sqrt{(T_1-t)}}{\sqrt{(Y_1-y_0)}} -\frac{\sqrt{(T_2-t)}}{\sqrt{(Y_2-y_0)}} =0.
\end{eqnarray}
It may be solved explicitly in the simplest case of the two hard scales set equal, $T_1=T_2=T$, and at $Y_1\simeq Y_2\simeq Y=\ln(1/x)$, i.e., in the $z$ region where $\ln(1/z)<<\ln(1/x)$ and $\ln(1/(1-z))<<\ln(1/x)$~\footnote{In spite of the large nonexponential factors like $\ln(1/x)$ [due to the singularity of the splitting function $P_{g \to gg}(z)$] the contribution from the integration region near the kinematical boundaries $z\sim x$ and $1-z \sim x$ is not dominant, since in this case the obtained exponential factor $\exp{[\sqrt{8N_c} \sqrt{Y(T-\tau)}]}$ is not leading [less than in Eq. (\ref{leading})]. }.
Then the saddle point is equal to
\begin{eqnarray}
y_0= Yt/(4T-3t)
\end{eqnarray}
and Eq.~(\ref{splitting-gluon-t}) reduces to
\begin{eqnarray}
\label{gluon-saddle} 
& x^2 D^{gg}_{h2}(x,x;\tau,T,T)\nonumber\\
& \sim \int\limits_{\tau}^{T}dt \int\limits_x^{1-x} dz
P_{g \to gg}(z) \exp{[\sqrt{8N_c} \sqrt{Y(4T-3t)}]}.
\end{eqnarray}
The $t$ integration is not a saddle-point type, and therefore, one of the edges, namely $t\to \tau$, dominates. That is,
\begin{equation}
\label{leading}
x^2 D^{gg}_{h2}(x,x;\tau,T,T)\sim \exp{[\sqrt{8N_c} \sqrt{Y(4T-3\tau)}]}.
\end{equation}

\section{\label{sec4}Discussion and Conclusions}
Now let us discuss in more detail what follows from our estimation of splitting integrals in the double logarithm approximation by the saddle-point method~\footnote{As in the usual DGLAP approach we consider only the evolution at not too small scales, larger than $Q_0>>\Lambda$. All contributions below $Q_0$ should be collected in the ``input two-parton distributions,'' which should already be  integrated over, $q^2<Q^2_0$.}.

For  single splitting diagrams ($1\times2$ contribution) the lower limit for the $t$ integration in the estimation~(\ref{gluon-saddle}) may be taken at the reference scale, i.e., $\tau=t(q^2)|_{q=\mu}=0$, due to the strong suppression factor $F_{2g}^2(q)$. The characteristic value of $q$ is of the order of the ``effective gluon mass'' $m_g\sim \mu$ in the further $q$ integration. Thus, one obtains for this contribution the following estimation:
\begin{eqnarray}
\label{saddle1x2} 
x^2 D^{gg}_{h2}(x,x;0,T,T) \sim
\exp{[\sqrt{8N_c} (\sqrt{YT}+\sqrt{YT})]}.
\end{eqnarray}
This means that the splitting takes place at the ``characteristic point'' with the scale $k^2$ close to $\mu^2$
and with the longitudinal momentum fraction $u\sim 1$ (the saddle point $y_0\sim t\sim \tau \sim 0$ in this case). After splitting, one has  two independent ladders with the well-developed  BFKL~\cite{ryskin,bfkl,bfkl2,bal} and DGLAP~\cite{gribov,lipatov,dokshitzer,altarelli} evolutions. Every ladder contributes to the cross section with the large exponential factor $\exp{[\sqrt{8N_c} \sqrt{YT}]}$, which is just the same as for single distributions [compare Eq.~(\ref{saddle1x2}) with Eq.~(\ref{mellin in x=0})]. Therefore, in the double logarithm approximation  single splitting diagrams~(\ref{splitting-gluon}) have, in fact, the factorization property if one takes only the leading exponential factors into consideration. However, the contributions to the cross section from single splitting diagrams and from the factorization component differ in nonexponential factors omitted here, especially, in the different ``normalization'' at the reference scale: {\it one} ``nonperturbative''  parton for single splitting diagrams and {\it two} initial independent ``nonperturbative''  partons for the factorization component.

The factorization property for the integral $D^{ij}_{h2}(x_1,x_2;\mu^2,Q_1^2,Q_2^2)$~(\ref{splitting}) was found also in Ref.~\cite{Snigirev:2010ds} in the double logarithm limit based on other techniques.

For  double splitting ($2\times2$) diagrams the leading exponential contribution arises from the lower 
limits of $t$ integration and either lower or upper limits of $q$ integrations, depending on the available rapidity interval $Y$. There is competition between the exponential factor caused by the evolution,
which prefers a small $\tau$, and the phase space factor in the $q^2$ integral. Because of the nonlogarithmic character of the  integration over $d^2q$ for a not sufficiently large $Y$, the contribution from the upper limit of $q$ may dominate. Indeed, let us consider the production of two $b\bar b$ pairs in a central rapidity ($\eta\sim 0$) region. That is, we take $T_1=T_2=T$,  $Y_1=Y_2=Y$ and keep just the leading exponential factors in the double parton distributions,
\begin{eqnarray}
\label{d2}
& & x^2 D_{h2}(x,x,q^2,Q^2,Q^2)\nonumber\\
& & \sim\exp(\sqrt{8N_cY(4T-3\tau)}-
2aT+a\tau)\, ,
\end{eqnarray}
where for better accuracy we keep the term $-at$ in the exponent of Eq.~(\ref{mellin in x=0}) [recall that here $t(q^2)=\tau$].

Thus the logarithmic $dq^2/q^2$ integral takes the form
 \begin{equation}
\label{l-int}
\int\frac{dq^2}{q^2}\exp(2\sqrt{8N_cY(4T-3\tau)}-4aT +2a\tau)q^2\, ,
\end{equation}
with  $\ln(q^2/\Lambda^2)=L=\ln(\mu^2/\Lambda^2)
e^{\beta\tau/2}$. The $L$ behavior of the integrand of Eq.~(\ref{l-int}),
%$$ f(L)=$$ 
 \begin{eqnarray}
\label{fL}
f(L)=& &\exp\left(2\sqrt{8N_cY(4T-3\tau)}-4aT+2a\tau
\right)\nonumber\\
& &\times\exp\left(\ln\left(\frac{\mu^2}{\Lambda^2}\right)
e^{\beta\tau/2}\right),
\end{eqnarray}
is shown in Fig.~1 in the case of $Y=5$ and $Y=2$, corresponding to the LHC energy $\sqrt s= 14$ TeV and the RHIC energy $\sqrt s= 500$ GeV (to be more or less realistic, here $Y$ is calculated as $Y=\ln(x_0/x)$ with $x_0\sim 0.2$). For this numerical estimation we take $N_c=3$, $n_f=4$ and $\Lambda=150$ MeV~\footnote{These parameters provide reasonable values of QCD coupling in a relevant region, say, $\alpha_s=0.29$ and $\alpha_s=0.22$ at $Q^2=4$ and 20 GeV$^2$, correspondingly.}; $Q^2=50$ GeV$^2$. In the LHC case we consider also the DPS $W$-boson production, taking $Q^2=10^4$ GeV$^2$
and $Y=3$.\\
\begin{figure}
\includegraphics[width=0.45\textwidth]{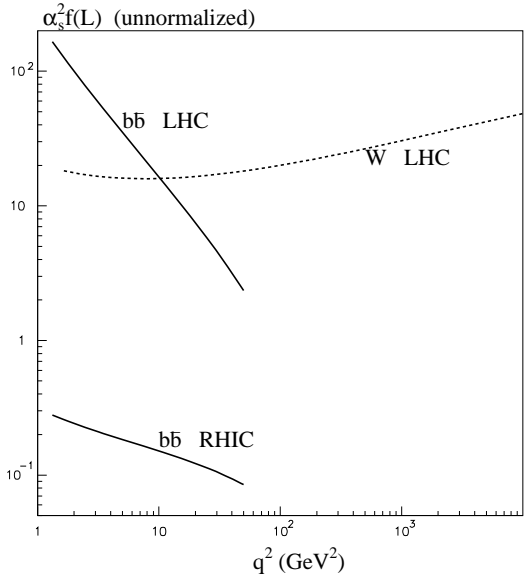}
\caption{ The $q$ dependence of the integrand $f(L)$ in the logarithmic scale.
\label{fig:figa}}
\end{figure}

As it is seen in Fig.~1, where the relevant quantity $\alpha_s^2(q^2)f(L)$ is plotted, for the DPS production of two $b\bar b$ pairs the major contribution comes from a low $q^2$. That is,  the reaction may be  effectively described by the $1\times 1$ term; the formation of two parton branches (one to two splitting) takes place mainly at low scales. However, at the RHIC energy, when the available rapidity interval
is not large, the $q^2$ dependence is not steep and  the contribution caused by the splitting somewhere in the middle of the evolution is still not negligible. The same can be said about the DPS $W$-boson production at the LHC. Here the upper edge of the $q^2$ integral dominates. This part may be described as the collision of {\it one} pair of partons supplemented by a more complicated, $2\to 4$ or $2\to 2W$, hard matrix element. However, clearly we need to also account for  contributions from the whole $q^2$ interval.

In other words, depending on the precise kinematics, we may deal either with a single parton pair collision (times the $2\to 4$ hard subprocess), with the contribution of the $1\times 1$ type, where the formation of two parton branches (one to two splitting) takes place at low scales, or with the $2\times 2$ configuration where the splitting may happen everywhere (with more or less equal probabilities) during the evolution. Note that just this last possibility may be relevant for the LHC experiments.

When both splittings take place at relatively small scales $\sim \mu^2<<Q^2$, the $2\times2$ contribution {\it cannot} be  considered as the result of the interaction of {\it one} pair of partons with the $2\to 4$ hard subprocess, unlike the statement in Refs.~\cite{stir,Blok:2011bu,Gaunt:2012wv,Manohar}.

Recall also that there may be a configuration with two quite different scales (say, $Q^2_1<<Q^2_2$), in which the upper limit of the $q^2$ integral is given by a smaller scale (at $q>Q_1$ the hard matrix element corresponding to $\sigma^A$ begins to diminish with $q_t$). In this case  the collinear evolution from the scale $q=Q_1$ up to the scale $Q_2$  in the ladders is sufficiently justified. A configuration with two quite different ``final'' rapidities [for instance, $1<< \ln{(1/x_1)}<<\ln{(1/x_2)}$] is also interesting to probe single and double splitting diagrams, since the BFKL evolution takes place in the ladders before and after (in one of ladders only) splitting. 

Here it is worth noticing that the asymptotic prediction mainly ``teaches'' us a tendency and tells us nothing practical about values of $x_1$, $x_2$, $Q_1$, and $Q_2$, beginning from which the asymptotic behavior is a good approximation to the real one.
Therefore, it makes sense to consider the quantitative contribution of the $2\times 2$ term even within the collinear approach as applied to the LHC kinematics, where the large available values of $Q_1$ and $Q_2$ (in comparison with $m_g$ and $\mu$), $\ln{(1/x_1)}$ and $\ln{(1/x_2)}$ (in comparison with 1) can provide  configurations with the BFKL or DGLAP evolution in ladders before and after splitting, depending on the processes under consideration.

A number of processes were suggested in order to probe DPS at the LHC. Promising candidate processes, such as same-sign $W$ production, $Z$ production in association with jets, four-jet production, production of a $b {\bar b}$ pair with two jets, production of a $b{\bar b}$ pair with $W$, have been discussed in detail in a review~\cite{Bartalini:2011jp}, with many references to prior works listed therein.
Quite recently the processes with pairs of heavy quarkonia in the final state were considered~\cite{kom,Baranov:2011ch,Novoselov} as precise probes of the DPS at the LHC.
We believe that production of a $b{\bar b}$ pair (or $J/\psi$) with $W$ may be a good candidate process to probe the QCD evolution of the double distribution functions due to a configuration with two quite different scales.

For completeness, it is also interesting to estimate the value of the exponential factors available at the LHC kinematics in single ladder diagrams. The asymptotic behavior of the distribution functions is determined by the factor 
\begin{eqnarray}
\label{exponet} 
\exp{[\sqrt{8N_c} \sqrt{YT}]}=\exp{[2.4\sqrt{k_{\rm BFKL}} \sqrt{k_{\rm DGLAP}}]},
\end{eqnarray}
where 
$$k_{\rm BFKL}=Y=\ln{(1/x)}< 14 $$
at $ x_{\rm min}=10^{-6},$ and

$$k_{\rm DGLAP}=\ln{[(\ln{(Q/\Lambda)}/(\ln{(\mu/\Lambda)}]}<1.4$$  
at $Q_{\rm max}=250$ GeV, $\mu =1$ GeV, $\Lambda =0.15$ GeV. In  Eq.~(\ref{exponet}) we put $n_f=4, N_c=3$. 

However, it is better to compare the pure BFKL factor 

$\omega_0\ln(1/x)\sim (\alpha_s N_c/\pi)\ln(1/x)$ with the analogous DGLAP factor $2N_ct$. In the BFKL case the resummation of the next-to-leading logarithmic corrections leads to $\omega_0\sim 0.3$~\cite{bfkl3}, that is, the BFKL power $\omega_0*\ln(1/x)<4$, while for the DGLAP evolution we have $(4N_c/\beta)k_{\rm DGLAP}<2$.
Thus the LHC kinematics admits a wider interval for the BFKL evolution than for the DGLAP one in compatible dimensionless variables. Therefore, it may be interesting (and even more justified theoretically) to consider the multiple parton interactions in the framework of the BFKL approach (see, for instance, Ref.~\cite{flesburg}). \\

In summary, we have demonstrated that all components of the generalized double distribution functions have the factorization structure in the double logarithm approximation and contribute to the cross section with the same leading exponential terms in $Y$ and $T$, but with different weights (nonexponential factors). For the debatable double splitting diagrams, depending on the precise kinematics, we may deal either with a single parton pair collision (times the $2\to 4$ hard subprocess), with the contribution of the $1\times 1$ type where the formation of two parton branches (one to two splitting) takes place at low scales, or with the $2\times 2$ configuration where the splitting may happen everywhere (with more or less equal probabilities) during the evolution. In order to probe the QCD evolution of the double distribution functions better, we suggest to also investigate the processes with two quite different scales, in particular, production of a $b{\bar b}$ pair (or $J/\psi$) with $W$, which was estimated at the LHC kinematical conditions in Ref.~\cite{Berger:2011ep} using the factorized component only.
\begin{acknowledgments}
Discussions with A.I.~Demianov and N.P.~Zotov are gratefully acknowledged. This work is partly supported by Russian Foundation for Basic Research Grants  No. 10-02-93118 and No 12-02-91505 and by the Federal Program of the Russian State RSGSS=65751.2010.2. 
\end{acknowledgments}

%\clearpage


\begin{thebibliography}{99}
\bibitem{Bartalini:2011jp} P.~Bartalini {\it et al.,}  arXiv:1111.0469.
\bibitem{AFS} T.~Akesson {\it et al.} (AFS Collaboration), Z. Phys. C {\bf 34}, 163 (1987).
\bibitem{UA2} J.~Alitti {\it et al.} (UA2 Collaboration),  Phys. Lett. B {\bf 268}, 145 (1991).
\bibitem{cdf4jets} F.~Abe {\it et al.} (CDF Collaboration), Phys. Rev. D {\bf 47}, 4857 (1993).
\bibitem{cdf} F.~Abe {\it et al.} (CDF Collaboration), Phys. Rev. D {\bf 56}, 3811 (1997).
\bibitem{D0} V.M.~Abazov {\it et al.} (D0 Collaboration), Phys. Rev. D {\bf 81}, 052012 (2010).
\bibitem{atlas} ATLAS Collaboration, ATLAS-CONF-2011-160. 
\bibitem{DelFabbro:2002pw} A.~Del Fabbro and D.~Treleani, Phys.\ Rev.\  D {\bf 61}, 077502 (2000); {\bf 66}, 074012 (2002).
\bibitem{Hussein:2006xr} M.Y.~Hussein, Nucl.\ Phys. B (Proc.\ Suppl.)\  {\bf 174}, 55 (2007).
\bibitem{Kulesza:1999zh} A.~Kulesza and W.J.~Stirling, Phys.\ Lett.\  B {\bf 475}, 168 (2000).
\bibitem{Cattaruzza:2005nu} E.~Cattaruzza, A.~Del Fabbro and D.~Treleani, Phys.\ Rev.\  D {\bf 72}, 034022 (2005).
\bibitem{maina} E.~Maina, J. High Energy Phys.  04 (2009) 098; 09 (2009) 081.
\bibitem{Berger:2009cm}E.L.~Berger, C.B.~Jackson, and G.~Shaughnessy, Phys. Rev. D {\bf 81}, 014014 (2010).
\bibitem{Berger:2011ep}E.L.~Berger, C.B.~Jackson, S. Quackenbush, and G.~Shaughnessy, Phys. Rev. D {\bf 84}, 074021 (2011).
\bibitem{Diehl:2011yj} M.~Diehl, D.~Ostermeier, and A.~Schafer, J. High Energy Phys. 03 (2012) 089.
\bibitem{Ryskin:2011kk} M.G. Ryskin and A.M. Snigirev, Phys. Rev. D {\bf 83}, 114047 (2011).
\bibitem{Blok:2010ge}B.~Blok, Yu.~Dokshitzer, L.~Frankfurt, and M.~Strikman, Phys. Rev. D {\bf 83}, 071501 (2011).
\bibitem{Diehl:2011tt} M.~Diehl and A.~Schafer, Phys. Lett. B {\bf 698}, 389 (2011).
\bibitem{stir} J.R. Gaunt and W.J. Stirling, J. High Energy Phys.  06 (2011) 048.
\bibitem{flesburg} C.~Flensburg, G. Gustafson, L. Lonnblad, and A. Ster, J. High Energy Phys. 06 (2011) 066.
\bibitem{Kirschner:1979im} R.~Kirschner, Phys.\ Lett.\  B {\bf 84}, 266 (1979).
\bibitem{Shelest:1982dg} V.P.~Shelest, A.M.~Snigirev, and G.M.~Zinovjev, Phys.\ Lett.\  B {\bf 113}, 325 (1982); Teor. Mat. Fiz. {\bf 51}, 317 (1982) [Theor. Math. Phys. {\bf 51}, 523 (1982)].
\bibitem{snig03} A.M.~Snigirev, Phys. Rev. D {\bf 68}, 114012 (2003).
\bibitem{snig04} V.L.~Korotkikh and A.M.~Snigirev, Phys. Lett. B {\bf 594}, 171 (2004).
%\bibitem{snig08} A.M.~Snigirev, arXiv:0809.4377 [hep-ph].
\bibitem{gribov} V.N.~Gribov and L.N.~Lipatov, Yad. Fiz. {\bf 15}, 781 (1972) [Sov. J. Nucl. Phys. {\bf 15}, 438 (1972)]; Yad. Fiz. {\bf 15}, 1218 (1972) [Sov. J. Nucl. Phys. {\bf 15}, 675 (1972)].
\bibitem{lipatov} L.N.~Lipatov, Yad. Fiz. {\bf 20}, 181 (1974) [Sov. J. Nucl. Phys. {\bf 20}, 94 (1974)].
\bibitem{dokshitzer} Yu.L.~Dokshitzer, Zh. Eksp. Teor. Fiz. {\bf 73}, 1216 (1977) [Sov. Phys. JETP {\bf 46}, 641 (1977)].
\bibitem{altarelli} G.~Altarelli and G.~Parisi, Nucl. Phys. B {\bf 126}, 298 (1977). 
\bibitem{Blok:2011bu} B.~Blok, Yu.~Dokshitzer, L.~Frankfurt, and M.~Strikman, Eur. Phys. J. C {\bf 72}, 1963 (2012).
\bibitem{Gaunt:2012wv} J.R. Gaunt and W.J. Stirling, arXiv:1202.3056 [hep-ph].
\bibitem{Frankfurt:2002ka} L.~Frankfurt and  M.~Strikman, Phys. Rev. D {\bf 66}, 031502 (2002).
\bibitem{Manohar} A.V. Manohar and W.J. Waalewijn, Phys. Lett. B {\bf 713}, 196 (2012).
\bibitem{ryskin}L.V.~Gribov, E.M.~Levin, and M.G.~Ryskin, Phys. Rep. {\bf 100}, 1 (1983).
\bibitem{bfkl}E.A.~Kuraev, L.N.~Lipatov, and V.S.~Fadin, Zh. Eksp. Teor. Fiz. {\bf 71}, 840 (1976) [Sov. Phys. JETP {\bf 44}, 443 (1976)].
\bibitem{bfkl2}E.A.~Kuraev, L.N.~Lipatov, and V.S.~Fadin, Zh. Eksp. Teor. Fiz. {\bf 72}, 377 (1977) [Sov. Phys. JETP {\bf 45}, 199 (1977)].
\bibitem{bal}I.I.~Balitsky and L.N.~Lipatov, Yad. Fiz. {\bf 28}, 1597 (1978) [Sov. J. Nucl. Phys. {\bf 28}, 822 (1978)].
\bibitem{cteq} J. Pumplin {\it et al.,} J. High Energy Phys.  07 (2002) 012.
\bibitem{Snigirev:2010ds} A.M.~Snigirev, Phys. Rev. D {\bf 83}, 034028 (2011).
\bibitem{kom} C.-H. Kom, A. Kulesza, and W.J. Stirling, Phys. Rev. Lett. {\bf 107} 082002 (2011).
\bibitem{Baranov:2011ch} S.P. Baranov, A.M. Snigirev, and N.P. Zotov, Phys. Lett. B {\bf 705} 116 (2011).
\bibitem{Novoselov} A.A. Novoselov, arXiv:1106.2184 [hep-ph].
\bibitem{bfkl3} M. Ciafaloni, D. Colferai, and G. Salam, Phys. Rev. D {\bf 60}, 114036 (1999); G. Salam, JHEP {\bf 9807}, 019 (1998); Act. Phys. Pol. B {\bf 30}, 3679 (1999);
V.A. Khoze, A.D. Martin, M.G. Ryskin, and W.J. Stirling, Phys. Rev. D {\bf 70}, 074013 (2004).
\end{thebibliography}
\end{document}